# Origin of Low-Frequency Negative Transconductance Dispersion in p-HEMT's

**V. R. Balakrishnan**
Solid State Physics Laboratory,
Lucknow Road, Timarpur, Delhi-110054, India

**Vikram Kumar**
National Physical Laboratory
New Delhi-110012, India

**Subhasis Ghosh**
School of Physical Sciences,
Jawaharlal Nehru University,
New Delhi-110067, India

**Abstract**

Measurements of low-frequency transconductance dispersion at different temperatures and conductance deep level transient spectroscopic(CDLTS) studies of an AlGaAs/InGaAs pseudomorphic HEMT were carried out. The experimental results show the presence of defect states at the AlGaAs/InGaAs hetero-interface. A mobility degradation model was developed to explain the low frequency negative transconductance dispersion as well as the apparent 'hole' like peaks observed in the CDLTS spectra. This model incorporates a time dependent change in 2DEG mobility due to ionised impurity scattering by the remaining charge states at the adjoining AlGaAs/InGaAs hetero-interface.

Indexing Terms- Deep Level Transient Spectroscopy, interface states, pseudomorphic HEMT, mobility degradation, Transconductance dispersion.

# I    Introduction

AlGaAs/InGaAs pseudomorphic high electron mobility transistors(p-HEMT's) exhibit high transconductance due to carrier confinement, which makes them very attractive for high frequency applications. The high transconductance is susceptible to surface and interface related degradation effects such as the low-frequency dispersion of transconductance[1,2]. The physical origin and the location of the defect states causing this dispersion has not yet been clearly established. Defect states can be distributed within different contacts, layers, surfaces and interfaces between dissimilar materials whose contributions cannot be easily separated or determined. The likely defect states in each of the layers and interfaces of a HEMT structure such as used in our study, shown in Fig. 1, are (i)surface states in the ungated regions[3,4], (ii)DX centres in the top GaAlAs layer[5], and (iii)interface states between dissimilar layers[6].

Surface states in the ungated regions between the gate and source/drain contacts have been reported to cause the negative transconductance dispersion in metal-semiconductor field effect transistors(MESFET's)[4] and HEMT's[1]. On the other hand, positive transconductance dispersion has been observed in GaAlAs/GaAs HEMT, which has been attributed to bulk traps such as DX centres and intrinsic point defects present in the GaAlAs layer[1]. There is strong evidence that the negative $g_m$ dispersion and apparent 'hole' like peaks in the conductance deep level transient spectroscopy(CDLTS) spectra[7] are correlated and are due to trapping and detrapping of carriers from surface states in the ungated regions in the case of MESFET's[4].

Deep level traps and recombination centres at interfaces of heterostructures such as GaAlAs/GaAs and GaInAs/GaAs have also been found to degrade the performance of devices such as DH lasers[8]. The interfacial properties of the heterostructure strongly depend on the growth technique such as molecular beam epitaxy (MBE), and its associated growth parameters. The interface states are generally present as localized states in both the materials near the interface[9]. These states can cause additional scattering of electrons confined in the quantum well by trapping and tunnelling processes and may be one of the sources of low frequency noise in HEMT's[10].

In the present work, CDLTS and transconductance measurements at different frequencies and temperatures have been carried out. A model has been proposed to explain the observed dispersion. It assumes the presence of defect states at the AlGaAs/InGaAs interface close to the two-dimensional electron gas (2DEG). The charge present in the defect layer will be shown to cause a time dependent mobility degradation resulting from ionised impurity scattering.

## II  Experimental

The device used in this work is a commercial GaAlAs/InGaAs/GaAs pseudomorphic HEMT, procured from NEC, Japan. The typical transconductance of the device is 60mS at $V_{DS}$ =2V, $I_{DS}$ =10 mA[11]. The HEMT structure is grown by metal organic chemical vapour deposition(MOCVD) technique. The gate metal used is Ti/Al and the dimensions of the gate are 0.2μm by 200 μm. Capacitance voltage profiling at different temperatures was carried out using the Boonton 72B capacitance meter(1MHz), HP4140B precision voltage source and a liquid nitrogen cryostat. The frequency dependence of AC transconductance on applied gate voltage was studied using the current amplifier, HP8116A function generator and the SRS 830DSP lock-in amplifier. The conductance DLTS spectra were obtained using Polaron DL4600 DLTS system along with the current amplifier.

## III  Experimental Results

Capacitance-Voltage measurements on the gate Schottky contact were carried out to estimate the maximum 2DEG density $n_{so}$ using the linear charge control model[12]. Ionised impurity concentration versus depth and applied voltage were extracted from the C-V measurements as shown in Fig. 2(a) to locate the position of the quantum well with respect to the surface[13]. AC transconductance measurements at different gate voltages were performed by applying a small ac signal of 10mV over a quiescent gate bias $V_{DS}$ =20mV. The transconductance $g_m$ has a peak around DC gate bias of –0.25 V that coincides with the peak in the carrier concentration as shown in Fig.2(a). The figure also shows a frequency dispersion of $g_m$ at frequencies between 10Hz and 10KHz. This dispersion is much more evident in Fig.2(b) where the dispersion in $g_m$ at different frequencies with respect to the measured $g_m$ at 10Hz is plotted. This figure shows a negative dispersion with a maximum near 1KHz. Moreover, the dispersion also peaks around gate voltages of -0.25V. The depth of the depletion layer edge around this reverse gate voltage approximately corresponds to the spatial position of the quantum well as shown in Fig.2(a). Similarly, at a voltage of +0.2V corresponding to the surface little dispersion is seen. This observation contrasts with similar measurements carried out on MESFET's[14] where the transconductance dispersion is much more at a gate bias of 0V compared to dispersion near the peak in the transconductance due to surface states. We have shown similar behaviour[15] in GaAs-based MESFET's. Therefore we conclude that the cause of transconductance dispersion in this study is not due to surface states present in the ungated regions but instead caused by interface states present near the quantum well. Fig. 3 shows the CDLTS spectra where a apparent 'hole' like peak is observed. The spectra shows the presence of two traps which get more resolved at lower rate windows. However, due to the nature of the spectra, it was possible to estimate activation energy $E_T \approx 400$meV and capture cross section $\sigma_\infty$ in the range of $10^{-15}$cm$^2$ only for the low temperature peak. The measurement of transconductance using the lock-in technique was performed by applying a small sinusoidal signal of 10mV at the gate with dc bias $V_{GS}$=0. The source-to-drain current, $I_{DS}$, was measured under a small $V_{DS}$=20mV. The results showing the variation of transconductance with temperature at five different frequencies are depicted in Fig.4. These plots clearly indicate very strong frequency and temperature dependence. There appears a decrease followed by an increase, which shifts towards higher temperature with increasing frequency. The amplitude of the $g_m$ dispersion reduces at higher frequencies and ultimately disappears, as the interface traps are not able to respond to high frequency modulation.

**IV     Discussions**

Negative transconductance dispersion and 'hole' like peaks in CDLTS spectra have been attributed to presence of surface states in the ungated regions in devices such as MESFET's[3,4,7,14]. During the last decade, these unfavourable effects have been considerably lowered by reducing the interelectrode spacings[16] and lowering source and drain series resistances[17]. In addition, the incorporation of self aligned recessed gate structure with $n^+$ GaAs contact layer under the source and the drain have effectively eliminated the effect of surface states on transconductance dispersion.

The possibility of imperfections at the hetero-interfaces between ternary alloy compounds due to lattice mismatch, growth-front roughness caused by non-ideal surface kinetics during the MBE growth and clustering by compositional non-uniformity has been described by Hong et al.[18]. Higher noise levels in InGaP/GaAs HEMT's compared to lattice matched GaAlAs/GaAs HEMT's have been ascribed to scattering due to interface roughness resulting in degradation of low field mobility[9]. In the case of p-HEMT, the thickness of the high mobility InGaAs layer grown on GaAs is kept below a critical thickness to produce a dislocation free interface. Although the InGaAs layer may be dislocation free, the further growth of the electron donating GaAlAs layer can breakdown the translational symmetry at the interface resulting in localized trap centers in the vicinity of the GaAlAs/InGaAs heterointerface. These states behave as electron traps and affect the mobility of electrons in the 2DEG[6]. The reduction in the mobility due to the presence of interface traps has been ascribed to ionised impurity scattering. This is caused by the remaining charge state of the traps after the electrons are emitted when the reverse bias pulse is applied to the gate[6]. To develop a model to explain negative 'hole' like peak caused by the decrease in the 2DEG mobility, we assume that the traps have an energy level of $E_T$ measured from the bottom of the GaInAs conduction band. We also assume that the GaInAs conduction band remains unchanged with bias applied at the gate[19]. We shall further assume that the traps are present in a thin interfacial layer of thickness $\delta$ with a trap concentration of $N_{IS}$ /cm$^2$. Taking $q$ as electronic charge, $Z$ as gate width and $L$ as gate length, the change in the time dependent drain current $I'_{DS}$ during the emission CDLTS mode will be given by

$$I'_{DS}(t) = q\left(\frac{Z}{L}\right) V_{DS} \mu'_{2D}(t) n'_s(t) \quad . \tag{1}$$

Since the 2DEG mobility depends inversely on the ionised impurity charge in the case of ionized impurity scattering[20], the 2DEG mobility will be given by $\mu_{2D} \propto 1/N_D$ and instantaneous mobility in the presence of interface traps will be $\mu_{2D}'(t) \propto 1/[N_D+N_t(t)]$. Taking $N_t(t) = (N_{IS}/\delta)(1-exp(-e_n t))$, the instantaneous mobility can be written as

$$\mu'_{2D} = \frac{\mu_{2D}}{1 + \frac{N_{IS}}{N_D \delta}[1-\exp(-e_n t)]} \quad , \tag{2}$$

where $N_D$ is the net donor concentration in the InGaAs layer and $e_n$ is the emission rate of the interface states. The instantaneous increase in the net 2DEG electron concentration n'$_s$ due to emission of carriers from the traps will be

$$n'_s(t) = n_s + N_{IS}[1-exp(-e_n t)] \quad . \tag{3}$$

Here $N_{IS}$ is the two-dimensional interface state density of the defect layer. The $N_{IS}$ for the heterostructure grown under optimum conditions is of the order of 1 x $10^{10}$/cm$^2$[14], which is much smaller than $n_s$ (1x$10^{12}$/cm$^2$). The second term in equation 3 can therefore be neglected with n'$_s$ ≈ n$_s$. Using binomial expansion assuming ($N_{IS}$/$N_D\delta$) exp(-e$_n$t) to be much smaller than 1, equation 2 becomes,

$$\mu'_{2D}(t) = \frac{\mu_{2D}\left(1 + \frac{N_{IS}}{N_D\delta}[1 + \exp(-e_n t)]\right)}{\left(1 + \frac{N_{IS}}{N_D\delta}\right)^2} \qquad (4)$$

Equation 1 can then be written as,

$$I'_{DS}(t) = \frac{I_{DS}\left(1 + \frac{N_{IS}}{N_D\delta}[1 + \exp(-e_n t)]\right)}{\left(1 + \frac{N_{IS}}{N_D\delta}\right)^2} \qquad (5)$$

This equation suggests that the normalised instantaneous drain current $I_{DS}$ will result in a positive transient giving a negative 'hole' like peak in CDLTS emission spectra. Such transients have also been obtained by Takikawa[6] by solving the Poisson equation and the Schroedinger equation self consistently.

The same model can be extended to explain the origin of transconductance dispersion. In the linear regime, where the applied drain to source voltage $V_{DS}$ is small, both the intrinsic transconductance $g_{m0}$ and the series resistance $R_s$ between the source and the gate will be functions of mobility μ and electron density n. In terms of angular frequency *ω=2πf* which is applied at the gate, the instantaneous 2DEG mobility in equation 2 becomes,

$$\mu'_{2D}(f,T) = \frac{\mu_{2D}}{1+D(f,T)} \qquad (6)$$

where $D(f,T)$ is the dimensionless dispersion term[3]

$$D(f,T) = \frac{N_{IS}}{N_D \delta} \exp\left(\frac{-e_n}{\omega}\right). \qquad (7)$$

The thermal emission rate $e_n$ is given by

$$e_n = N_c \sigma_n V_{th} \exp(-E_T/k_B T) \qquad (8)$$

Here, $N_c$ is density of states in the conduction band, $\sigma_n$ is the capture cross section and $V_{th}$ is the average thermal velocity of carriers. The series resistance $R_s$ will comprise of a parallel combination of three resistances due to the three parallel conduction paths of the GaAlAs supply layer, the InGaAs electron confinement layer and the GaAs buffer layer. The least resistance offered would obviously be from the InGaAs layer and thus would dominate the resistance $R_s$. Neglecting the other two contributions, the resistance $R_s$ at $V_{GS} = 0$ will be

$$R_s = \frac{L_{sg}}{qWn_s\mu_{2D}}, \qquad (9)$$

given by

where $L_{sg}$ is the source to gate spacing, W is the width of the gate and q being the electronic charge. The dispersion dependent intrinsic transconductance $g'_{m0}$ and the series resistance $R'_s$, after substituting for instantaneous mobility $\mu'_{2D}$ from equation 6, become

$$g'_{mo}(f,T) = \frac{g_{mo}}{1+D(f,T)} \qquad (10)$$

and

$$R'_s(f,T) = R_s(1 + D(f,T)) \quad . \tag{11}$$

The extrinsic transconductance $g'_m$ will then be

$$\frac{1}{g'_m(f,T)} = \frac{1}{g'_{mo}(f,T)} + R'_s(f,T) \quad . \tag{12}$$

Substituting for $g'_{mo}$ and $R'_s$ from equations 10 and 11, $g'_m$ becomes

$$g'_m(f,T) = \frac{g_{mo}}{[1 + D(f,T)](1 + R_s g_{mo})} \tag{13}$$

Expressing dispersion independent $g_{mo}$ in terms of $g_m(T)$ and substituting in equation 13, we obtain

$$g'_m(f,T) = \frac{g_m(T)}{1 + D(f,T)} \tag{14}$$

This equation indicates that the dispersion in $g_m$ is independent of the dispersion caused by series resistance $R_s$. The simulation results depicting the variation of $g_m$ vs. temperature at several frequencies using equation 14 are shown in Fig.5. This behaviour can be explained by the sudden or sharp transition in the dispersion term $D(f,T)$, caused by temperature dependent emission of electrons from the interface electron traps. This thermal emission starts when the term $e_n/\omega$ approaches 1 for a given temperature. The dispersion term $D(f,T)$ given by equation 7 suggests that the ratio $N_{IS}/N_D\delta$ will be most significant in determining the extent of $g_m$ dispersion. This simulation matches sufficiently with the experimental data depicted in Fig.4 to demonstrate the validity of the mobility degradation model. All the deep level parameters used were experimentally determined and are given in Table I. The magnitude of the term $N_{IS}/N_D\delta$ was used as a variable parameter in determining the extent of $g_m$ dispersion. The most likely value of the term $N_{IS}/N_D\delta$ was determined iteratively until a good match was obtained between the theoretical and experimental plots given in Fig.4. Assuming $\delta \sim 50\text{Å}$, and background doping of the InGaAs layer as $1 \times 10^{16} cm^{-3}$, the interface density $N_{IS}$ is obtained approximately as $1 \times 10^{10} cm^{-2}$.

## IV Conclusions

Trapping effects from the defect states at the AlGaAs/InGaAs interface of a p-HEMT by means of low frequency transconductance dispersion and CDLTS measurements have been experimentally examined. A mobility degradation model was developed and was successfully used to explain the cause of low frequency transconductance dispersion as well as the negative 'hole' like peaks in the CDLTS spectra. The change in the 2DEG mobility is attributed to ionised impurity scattering due to remaining charge pertaining to a rate dependent change in the occupancy of interface electron traps.

**Acknowledgments**

We thank Mr. NK Nayyar and Ms Rachna Thakur for their help and support during the experimental work. We are also grateful to Dr. R. Muralidharan and Mr. Anil Aggarwal for providing us with samples. We also thank Dr. Vikram Dhar and Ms. Rashmi for giving valuable suggestions while writing the manuscript.

**Table I**

**HEMT device dimensions, reported values, and experimentally determined parameters used for simulation of transconductance versus temperature plots.**

| SYMBOL | NOMENCLATURE | VALUE |
|---|---|---|
| $L_g$ | Gate Length | 0.20μm |
| W | Gate Width | 200μm |
| $V_{th}$ | Thermal velocity | $2 \times 10^7$ cm/sec |
| $n_s$ | 2DEG electron conc. | $1.06 \times 10^{12}$/cm$^2$ |
| $E_T$ | Activation energy | 400 meV |
| $\sigma_\infty$ | Capture cross section | $\sim 1 \times 10^{-15}$/cm$^2$ |
| $R_s$ | Source resistance | 0.25 Ohms |
| $N_{IS}/N_D\delta$ | Trap conc. Ratio | 0.2 |

**Figure Captions**

**Figure 1.** Schematic diagram of a pseudomorphic HJ-FET showing the location of possible defect centers.

**Figure 2.** **(a)** Measured ac transconductance at different frequencies as a function of applied gate voltage. The free carrier profile $N_{C-V}$ as a function of applied gate voltage is also shown. The position of the peak in the transconductance matches with the position of the quantum well determined from the C-V measurements. The AC signal applied at the gate is 10 mV with $V_{DS}$ of 20 mV. **(b)** Transconductance dispersion plotted with respect to $g_m$ at 10 Hz from Fig. 2(a).

**Figure 3.** CDLTS spectra showing negative 'hole' like peaks at different emission rates. The voltages are V(fill) = +0.1 V, V(reverse) = -0.4 V. The fill pulse width is 10 ms. Arrhenius plot for determining the activation energy of the traps is shown in the inset.

**Figure 4.** Experimental transconductance versus temperature plots at different frequencies. The voltages are $V_{DS}$ = 20 mV and $V_{GS}$ = 0V.

**Figure 5.** Simulated transconductance versus temperature at different frequencies using equation 14. Parameters given in Table I have been used. This simulation matches sufficiently with the experimental data depicted in Fig.4 to demonstrate the validity of the mobility degradation model. The temperature dependence of $g_m(T)$ used in the simulation is also shown.

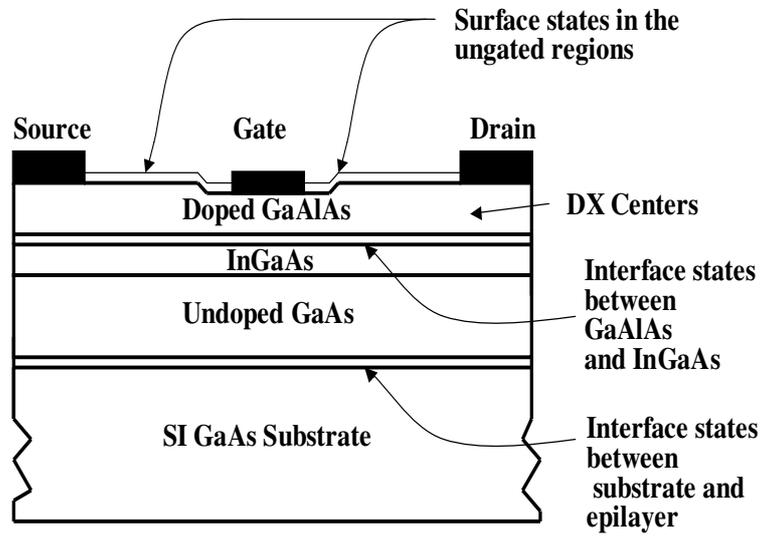

Figure 1.

V.R. Balakrishnan et al.

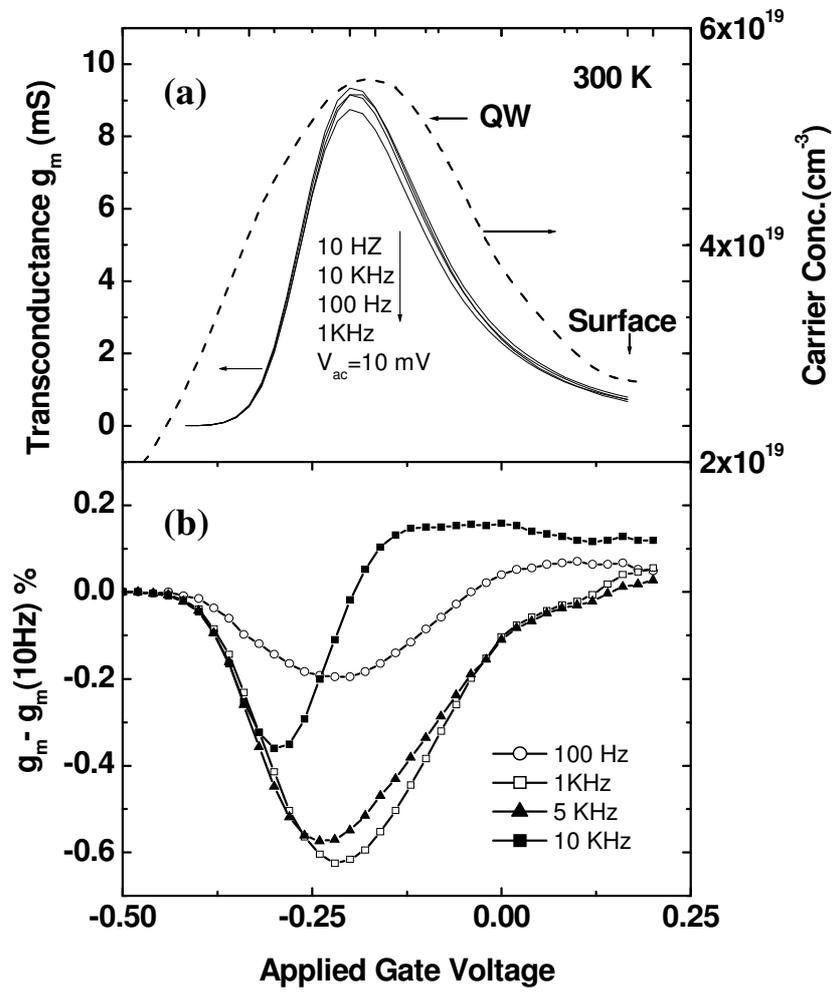

Figure 2

V.R. Balakrishnan et al.

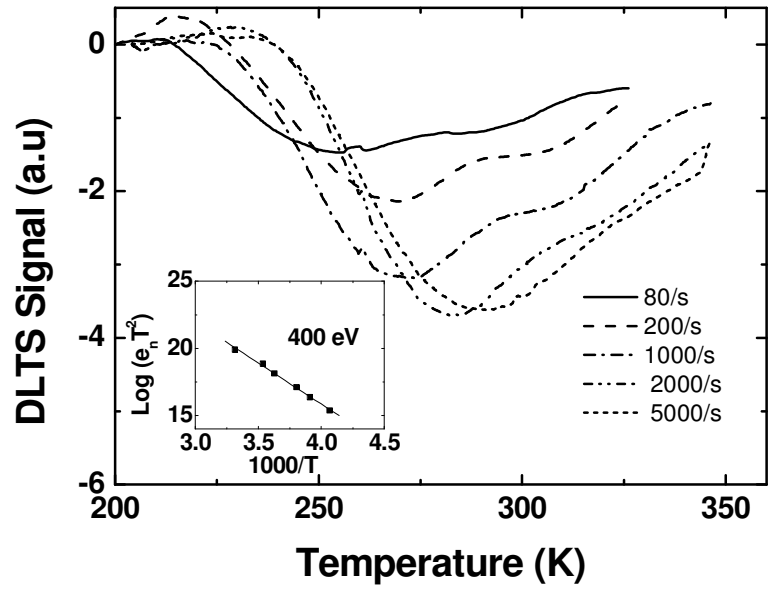

Figure 3.

V.R. Balakrishnan et al.

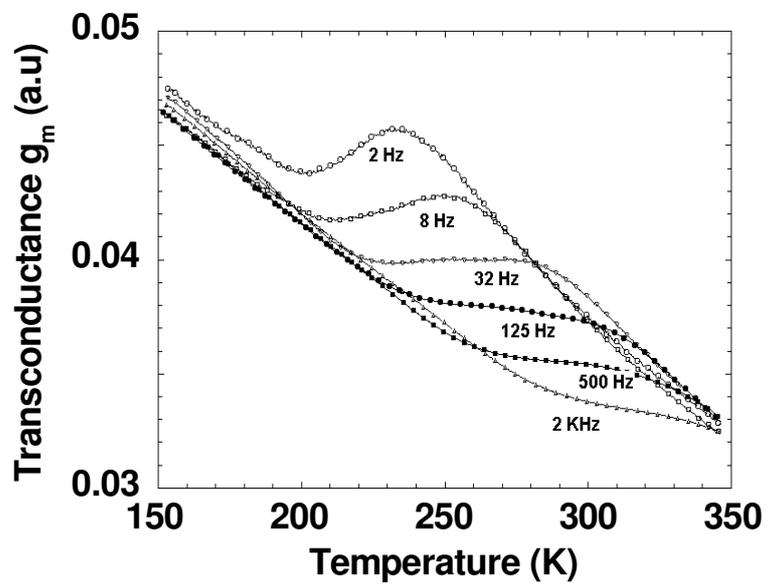

Figure 4.

V.R. Balakrishnan et al.

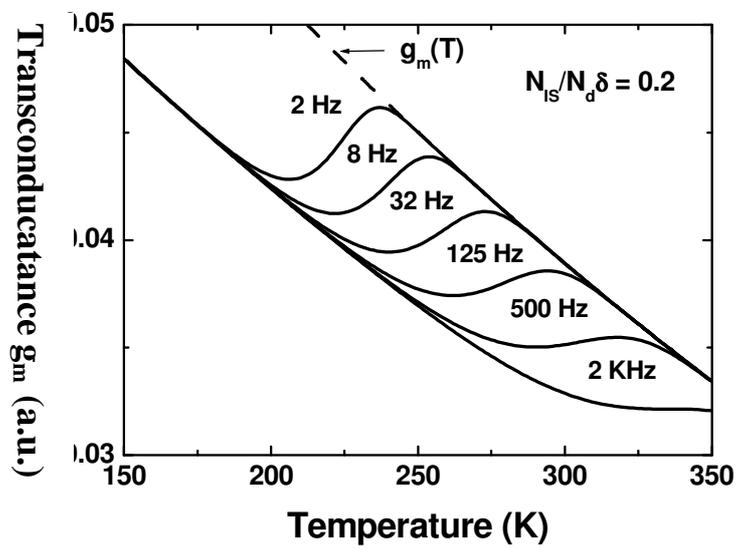

Figure 5.

V.R. Balakrishnan et al.